\documentclass[10pt,twoside]{article}

\usepackage{pst-all}
\usepackage{pstcol}
\usepackage{amssymb}
\usepackage[dvips]{graphicx}
\usepackage{amsmath}

\begin{document}

\title{Comparative kinetics of the snowball respect to other dynamical
objects}
\author{Rodolfo A. Diaz\thanks{%
radiaz@ciencias.unal.edu.co}, Diego L. Gonzalez\thanks{%
adrijara@tutopia.com}, Francisco Marin\thanks{%
pachomarin@lycos.com}, R. Martinez\thanks{%
romart@ciencias.unal.edu.co} \\
%EndAName
Universidad Nacional de Colombia.\\
Departamento de F\'{\i}sica. Bogot\'{a}, Colombia}
\maketitle

\begin{abstract}
We examine the kinetics of a snowball that is gaining mass while is rolling
downhill. This dynamical system combines rotational effects with effects
involving the variation of mass. In order to understand the consequences of
both effects we compare its behavior with the one of some objects in which
such effects are absent, so we compare the snowball with a ball with no mass
variation and with a skier with no mass variation nor rotational effects.
Environmental conditions are also included. We conclude that the comparative
velocity of the snowball respect to the other objects is particularly
sensitive to the hill profile and also depend on some retardation factors
such as the friction, the drag force, the rotation, and the increment of
mass (inertia). We emphasize that the increase of inertia could surprisingly
diminish the retardation effect owing to the drag force. Additionally, when
an exponential trajectory is assumed, the maximum velocity for the snowball
can be reached at an intermediate step of the trip.
\end{abstract}

\section{Introduction\label{sec:introduction}}

The snowball is a dynamical object that gains mass while is rolling
downhill. It is a particularly interesting problem since it permits to
combine several concepts of mechanics such as the traslational and
rotational dynamics, the rigid body, mass variable systems, normal and
tangential forces, radius of curvature for a trajectory etc \cite{Finn, mass
var}. Modeling this problem implies many input parameters and the use of
numerical methods. Additionally, an ansatz should be made about the way in
which the mass (or volume) is growing with time. Environmental conditions
are also considered utilizing well known assumptions about friction and drag
forces. The dynamical behavior of the snowball will be studied in the \emph{%
velocity} vs \emph{time}, and \emph{velocity} vs \emph{length} $-$planes.

Moreover, comparison with other dynamical objects could clarify many aspects
of the very complex behavior of the snowball. Therefore, we will develop our
analysis by comparing the snowball (\textbf{SB}) motion with the one
obtained from a skier sliding without friction (\textbf{SNF}), a skier
sliding with friction (\textbf{SF}) and a ball with constant mass and volume
(\textbf{B}).

In section \ref{sec:the problem}, we discuss the basic assumptions and write
out the equations of motion for the snowball. In section \ref{sec:comparison}%
, the comparison between the four dynamical objects mentioned above is
performed in the asymptotic regime. Section \ref{sec:consistency} describes
some proves of consistency to test our results. Section \ref{sec:analysis}
provides a complete analysis of the comparative kinetics of the four
dynamical objects based on some environmental retardation factors. Section %
\ref{sec:conclusions} is regarded for the conclusions.

\section{The problem of the snowball rolling downhill\label{sec:the problem}}

\subsection{Basic assumptions\label{sec:assumptions}}

The complete analysis of a snowball requires many input parameters. The
problem does not have any analytical solution, so it has to be solved
numerically. We have listed the numerical values assumed in this paper in
table \ref{tab:input} on page \pageref{tab:input}. Besides, we make the following assumptions in order
to get an equation of motion

\begin{enumerate}
\item The snowball is always spherical meanwhile is growing and acquiring
mass. Its density is constant as well.

\item It is supposed that there is only one point of contact between the
snowball and the ground, and that the snowball rolls without slipping
throughout the whole motion. We shall call it the No Slipping Condition (%
\textbf{NSC}).

\item In order to accomplish the previous condition, the frictional static
force that produces the rotation of the snowball ($F_{R_{S}}$), has to hold
the condition $F_{R_{S}}\leq \mu _{s}N\;$where $N\;$is the normal (contact)
force to the surface, and $\mu _{s}\;$is the coefficient of static friction.
We assume that $\mu _{s}\;$is independent on the load (Amontons' law), this
statement applies for not very large loads \cite{Pounder}.

\item The drag force owing to the wind is assumed to be of the form

\begin{equation}
F_{v}=-\frac{\rho _{A}C_{d}A}{2}v^{2}  \label{Drag force}
\end{equation}
where $\rho _{A}\;$is the air density, $C_{d}\;$is the air drag coefficient,
and $A\;$is the snowball's projected frontal area i.e. $A=\pi r^{2}.$

We assume the air drag coefficient $C_{d}\;$to be constant, since this
assumption has given reasonable results in similar problems \cite{Catalfamo}%
. On the other hand, it has been established that the force $F_{v}$ could be
a linear or quadratic function of the speed depending on the Reynolds number
($Re$) \cite{Parker}. For $Re>1\;$(which is our case) a quadratic dependence
fits well with experimental issues\cite{Shuttlecock}.

\item The mass of the snowball increases but finally reaches an asymptotic
value. Furthermore, a specific function of the mass (or volume) in terms of
time must be supposed. In our case we assume the following functional form 
\begin{equation}
M\left( t\right) =M_{0}+K_{0}\left( 1-e^{-\beta t}\right)
\label{Comport masa1}
\end{equation}

where $M_{0}\;$is the initial mass of the snowball and clearly the final
mass is $M_{0}+K_{0}.$

\item A hill profile must be chosen in order to make the simulation, like in
other similar problems \cite{Catalfamo}, it has an important effect on exit
speed. Specifically, we have chosen an inclined plane and an exponential
trajectory.
\end{enumerate}

\subsection{Equations of motion\label{sec:equations}}

\begin{center}
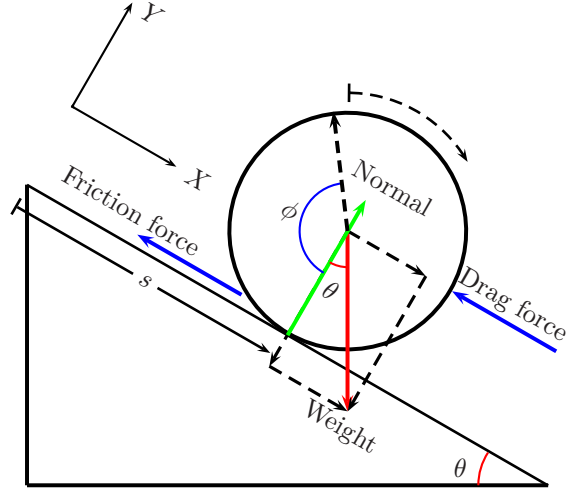
\begin{figure}[tbph]
%figure of the snowball on the wedge
\par
\begin{center}
\psset{unit=0.8cm} 
\begin{pspicture}(-0.5,1.5)(9,10.3)
\rput[br]{-30}(0,7)
{
%linea de base para posterior construcción del plano inclinado
\psline[linewidth=1pt](0,0)(10,0)
%la esfera
\pscircle[linewidth=1.5pt](5,2){2}
%su peso (W=3) inclinado 30 grados para posterior rotación
\rput[t]{30}(5,2){\psline[linewidth=1.5pt,linecolor=red]{->}(0,-3)}
%componente del peso en Y
\rput[t]{0}(5,2){\psline[linestyle=dashed,linewidth=1.2pt]{->}(0,-2.5981)}
\rput[t]{0}(6.5,2){\psline[linestyle=dashed,linewidth=1.2pt]{->}(0,-2.5981)}
%componente del peso en X
\rput[t]{0}(5,2){\psline[linestyle=dashed,linewidth=1.2pt]{->}(1.5,0)}
\rput[t]{0}(5,-0.59808){\psline[linestyle=dashed,linewidth=1.2pt]{->}(1.5,0)}
%fuerza de rozamiento
\psline[linewidth=1.5pt,linecolor=blue]{->}(4,0.2)(2,0.2)
%fuerza viscosa
\psline[linewidth=1.5pt,linecolor=blue]{<-}(7,2)(9,2)
%normal (N=3)
\psline[linewidth=1.5pt,linecolor=green]{->}(5,0)(5,2.5981)
%arcos que miden el ángulo del peso inclinado y el angulo barrido
\psarc[linecolor=red](5,2){0.6}{-90}{-60}
\psarcn[linecolor=blue](5,2){0.8}{-90}{-233.24}
%arco que determina el sentido de giro de la esfera
\psarcn[linewidth=1pt,linestyle=dashed]{|->}(5,2){2.3}{120}{60}
%vector que mide el ángulo barrido por la esfera en su rotación
\rput[t]{-143.24}(5,2){\psline[linewidth=1.5pt,linestyle=dashed]{->}(0,-2)}
%trazado de línea que demarca la distancia recorrida
\psline[linewidth=1.0pt]{|->}(0,-0.4)(4.9,-0.4)
%trazo de los ejes coordenados
\psaxes[labels=none,ticks=none]{->}(0,1.5)(0,1.5)(2,3.5)
%etiquetas
\rput*(2.5,-0.4){$s$}
\uput*[r](0,3.5){$Y$}
\uput*[r](2,1.5){$X$}
\rput[r](3,0.5){Friction force}
\rput[r](9,2.3){Drag force}
\rput[r](6,3){Normal}
\rput[r](7.2,-0.9){Weight}
\rput[c](5.25,1){$\theta$}
\rput[c](4.0,1.8){$\phi$}
}
%línea de la pared del plano inclinado
\psline[linewidth=1.5pt](0,7)(0,2)
%línea de la base del plano inclinado
\psline[linewidth=1.5pt](0,2)(8.6603,2)
%arco del ángulo del plano inclinado
\psarcn[linecolor=red](8.5,2){1}{180}{147}
%etiquetas
\rput[c](7.2,2.3){$\theta$}
\end{pspicture}
\end{center}
\caption{\textit{A snowball rolling downward on a wedge, the $X-$axis is
parallel to the wedge surface and the $Y-$axis is perpendicular.}}
\label{fig:bola1}
\end{figure}
\end{center}

To simplify the problem we start assuming the snowball rolling downhill on a
wedge whose angle is $\theta \;$(see Fig. \ref{fig:bola1}). In the time $t\;$%
the snowball has a mass $M\;$and in the time $t+dt\;$its mass is $M+dM$,$\;$%
let us consider a system consisting of the original snowball of mass $M\;$%
plus the little piece of snow with mass $dM$. At the time $t,\;$the momentum
of such system is $\mathbf{P}\left( t\right) =M\mathbf{v\;}$(bold letters
represent vectors) since the little piece of snow is still on the ground and
at rest. At the time $t+dt\;$the ball has absorbed the piece of snow
completely, so the momentum is given by $\mathbf{P}\left( t+dt\right)
=\left( M+dM\right) \left( \mathbf{v+}d\mathbf{v}\right) $ then the momentum
change is $d\mathbf{P=}$ $Md\mathbf{v}+\mathbf{v}dM\;$(where we have
neglected the differential of second order) and the total force will be 
\begin{equation}
\mathbf{F}=\frac{d\mathbf{P}}{dt}=M\frac{d\mathbf{v}}{dt}+\mathbf{v}\frac{dM%
}{dt}  \label{Fuerza variable}
\end{equation}%
where $\mathbf{v\;}$corresponds to the velocity of the center of mass
respect to the ground.

Further, the system is rotating too, we also suppose that such rotation is
always made around an axis which passes through the center of mass
(perpendicular to the sheet in Fig. \ref{fig:bola1}). In this case the
rotation is around a principal axis, hence the equation of motion for the
angular momentum is given by 
\begin{equation}
\mathbf{L}_{C}=I_{C}\overrightarrow{\omega }_{C}  \label{Mom ang C}
\end{equation}%
where the subscript $C\;$refers to the center of mass coordinates. $I_{C}\;$%
denotes the moment of inertia of the snowball measured from an axis passing
through the center of mass, and\ $\overrightarrow{\omega }_{C}\;$refers to
the angular velocity. According to figure \ref{fig:bola1}, $\mathbf{L}_{C}\;$%
is directed inside the sheet, and the torque will be

\begin{equation}
\frac{d\mathbf{L}_{C}}{dt}=\overrightarrow{\mathbb{\tau }}_{C}
\label{Newton rotat}
\end{equation}

We should remember that this equation is valid even if the center of mass is
not an inertial frame \cite{Finn}, which is clearly our case. To calculate $d%
\mathbf{L}_{C}$ we make an analogous procedure as in the case of $d\mathbf{P}
$, and the equation (\ref{Newton rotat}) is transformed into 
\begin{equation}
I_{C}\frac{d\overrightarrow{\mathbb{\omega }}_{C}}{dt}+\overrightarrow{%
\mathbb{\omega }}_{C}\frac{dI_{C}}{dt}=\overrightarrow{\mathbb{\tau }}_{C}\;.
\label{Torq variable}
\end{equation}
where $\overrightarrow{\mathbb{\tau }}_{C}\;$is the total torque measured
from the center of mass. For the sake of simplicity, we will omit the
subscript $C\;$from now on.

On the other hand, the external forces and torques exerted on the system are
similar to the ones in the simple problem of a ball on a wedge \cite{Finn} 
\begin{equation}
\mathbf{F}=\mathbf{W+N+F}_{Rs}+\mathbf{F}_{a};\;\overrightarrow{\mathbf{\tau 
}}=\mathbf{r\times F}_{Rs}  \label{Fuerza y torq}
\end{equation}%
where $\mathbf{W\;}$is the weight (which acts as the driving force), $%
\mathbf{N\;}$the normal force, $\mathbf{F}_{Rs}\;$the statical friction
force, and $\mathbf{F}_{a}\;$is any applied force which does not produce
torque. If we use as a coordinate system the one indicated in figure \ref%
{fig:bola1} (with the $Z-axis\;$perpendicular to the sheet) we can convert
these equations into scalar ones; then using Eqs. (\ref{Fuerza variable}), (%
\ref{Torq variable}) and (\ref{Fuerza y torq}) we get by supposing that $%
\mathbf{F}_{a}\;$is parallel to the $X-axis.$

\begin{eqnarray}
N-Mg\cos \theta &=&0\;,  \notag \\
Mg\sin \theta -F_{Rs}+F_{a} &=&M\frac{dv}{dt}+v\frac{dM}{dt}\;,  \notag \\
rF_{Rs} &=&I\frac{d\omega }{dt}+\omega \frac{dI}{dt}\;.  \label{Ec de mov}
\end{eqnarray}%
The first equation decouples from the others whenever the \textbf{NSC} is
maintained; so we will forget it by now. We should be careful in utilizing
the \textbf{NSC} since the radius is not constant. The correct \textbf{NSC}
in this case consists of the relation $ds=r\left( d\phi \right) \;$where $%
ds\;$is the differential length traveled by the center of mass of the
snowball (in a certain differential time interval $dt$),$\ d\phi \;$is the
angle swept along the differential path, and $r\;$is the snowball radius in
the time interval\ $\left[ t,t+dt\right] $. Using the correct \textbf{NSC},
we get

\begin{eqnarray}
v &=&\frac{ds}{dt}=r\frac{d\phi }{dt}=r\omega  \notag \\
&\Rightarrow &\omega =\frac{v}{r}  \label{NSC1}
\end{eqnarray}
and taking into account that the radius depends explicitly on the time, we
obtain 
\begin{equation}
\alpha =\frac{a}{r}-\frac{1}{r}\frac{dr}{dt}\omega  \label{NSC2}
\end{equation}
where $\omega \;$is the angular velocity,$\;\alpha \;$is the angular
acceleration,$\;$and $v,a\;$are the traslational velocity and acceleration
respectively\footnote{%
Observe that we can start from the traditional \textbf{NSC\ }with $v=\omega
r $. Nevertheless, the other traditional \textbf{NSC} $a=r\alpha $, is not
valid anymore.}.

It is convenient to write everything in terms of the displacement $s,\;$%
taking into account the following relations

\begin{equation}
v=\frac{ds}{dt}\;;\;a=\frac{d^{2}s}{dt^{2}}\;;\;\omega =\frac{d\phi }{dt}=%
\frac{1}{r}\frac{ds}{dt}  \label{NSC3}
\end{equation}
replacing (\ref{NSC1}, \ref{NSC2}, \ref{NSC3}) into (\ref{Ec de mov}), we
find

\begin{equation}
rF_{Rs}=I\left[ \frac{1}{r}\frac{d^{2}s}{dt^{2}}-\frac{1}{r^{2}}\frac{dr}{dt}%
\frac{ds}{dt}\right] +\left( \frac{1}{r}\frac{ds}{dt}\right) \frac{dI}{dt}
\label{Momento de inercia variable}
\end{equation}

Now we use the moment of inertia for the sphere $I=\left( 2/5\right)
Mr^{2}\; $and the fact the the Mass ``$M"\;$and the radius ``$r"\;$are
variable, then

\begin{eqnarray}
\frac{dI}{dt} &=&\frac{2}{5}\left( r^{2}\frac{dM}{dt}+2Mr\frac{dr}{dt}%
\right) \;,  \label{Derivada mom de inerc} \\
M &=&\frac{4}{3}\pi \rho r^{3}\;;\;\frac{dM}{dt}=4\pi \rho r^{2}\frac{dr}{dt}
\end{eqnarray}%
where the snowball density $\rho \;$has been taken constant. Additionally,
we assume the applied force $F_{a}\;$to be the drag force in Eq. (\ref{Drag
force}). With all these ingredients and replacing Eq. (\ref{Momento de
inercia variable}) into equations (\ref{Ec de mov}), they become 
\begin{equation*}
\frac{d^{2}s}{dt^{2}}+\frac{15\rho _{A}C_{d}}{56\rho }\frac{1}{r}\left( 
\frac{ds}{dt}\right) ^{2}+\frac{23}{7}\frac{1}{r}\frac{dr}{dt}\frac{ds}{dt}-%
\frac{5}{7}g\sin \theta =0\;,
\end{equation*}%
\begin{equation}
F_{Rs}=\frac{8}{3}\pi \rho r^{2}\left[ \frac{1}{5}r\left( \frac{d^{2}s}{%
dt^{2}}-\frac{1}{r}\frac{dr}{dt}\frac{ds}{dt}\right) +\left( \frac{ds}{dt}%
\right) \frac{dr}{dt}\right] \;.  \label{Ec final}
\end{equation}

Finally, to complete the implementation of Eqs. (\ref{Ec final}), we should
propose an specific behavior for $r\left( t\right) \;$or $M\left( t\right)
;\;$we shall assume that $M\left( t\right) \;$behaves like in Eq. (\ref%
{Comport masa1}). On the other hand, as a proof of consistency for Eqs. (\ref%
{Ec final}), we see that if we take $r=constant,\;$and $C_{d}=0,\;$we\ find
that

\begin{eqnarray}
\frac{d^{2}s}{dt^{2}} &=&\frac{5}{7}g\sin \theta  \notag \\
F_{Rs} &=&\frac{2}{7}M_{0}g\sin \theta  \label{Comport asint}
\end{eqnarray}
which coincides with the typical result obtained in all common texts for the
ball rolling downward on a wedge with the \textbf{NSC} \cite{Finn}.

\subsection{A snowball downhill on an arbitrary trajectory\label%
{sec:arbitrary}}

\begin{center}
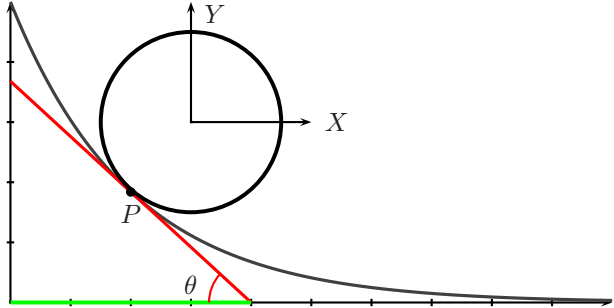
\begin{figure}[tbph]
\begin{center}
\psset{unit=0.8cm} 
\begin{pspicture}(0,0)(10,5)

\psaxes[labels=none,ticksize=1pt]{->}(0,0)(0,0)(10,5)
\psplot[plotstyle=curve,plotpoints=200,linecolor=darkgray,linewidth=1.2pt]{0}{10}{5 2.71828182 0.5 neg x mul exp mul}
\psplot[plotstyle=curve,plotpoints=200,linecolor=red,linewidth=1.2pt]{0}{4}{3.6788 0.91970 x mul sub}
\pscircle[linewidth=1.5pt](3,3){1.5320}
\psline[linewidth=1.5pt,linecolor=green](0,0)(4,0)
\psdots(2,1.8394)
\psarcn[linecolor=red](4,0){0.7}{180}{137.40}
\psaxes[labels=none,ticks=none]{->}(3,3)(3,3)(5,5)
%labels
\uput[d](2,1.8394){$P$}
\rput(3,0.3){$\theta$}
\uput[r](3,4.8){$Y$}
\uput[r](5,3){$X$}
\end{pspicture}
\end{center}
\par
%%%%
\caption{\textit{A snowball rolling downward on an exponential trajectory.
To find the local value of the angle $\protect\theta \;$we define the $%
X,Y-axis\;$as indicated in the figure. We see that for a sufficiently large
value of $x\;$we get $\protect\theta \rightarrow 0.$}}
\label{fig:bola2}
\end{figure}
\end{center}

In this case the acceleration written above converts into the tangential
one, and the equation (\ref{Ec de mov}) for the normal force becomes

\begin{equation}
N-Mg\cos \theta =M\frac{v^{2}}{R}  \label{Primer ec normal}
\end{equation}%
where $R\;$is the radius of curvature. In order to solve Eq. (\ref{Primer ec
normal}), it is convenient to use the coordinate axis plotted in figure \ref%
{fig:bola2}. In cartesian coordinates, the radius of curvature is given by

\begin{equation}
R\left( x\right) =\frac{\left[ 1+\left( y^{\prime }\right) ^{2}\right] ^{3/2}%
}{y^{\prime \prime }}\;;\;y^{\prime }\equiv \frac{dy}{dx}
\label{R de curvat}
\end{equation}
Moreover, the angle $\theta \;$is not constant any more, and according to
the figure \ref{fig:bola2} we see that

\begin{eqnarray}
\sin \theta &=&\frac{-dy}{ds}=\frac{-dy}{\sqrt{\left( dx\right) ^{2}+\left(
dy\right) ^{2}}}=-\frac{y^{\prime }}{\sqrt{\left[ 1+\left( y^{\prime
}\right) ^{2}\right] }}  \notag \\
\cos \theta &=&\frac{dx}{ds}=\frac{1}{\sqrt{\left[ 1+\left( y^{\prime
}\right) ^{2}\right] }}  \label{Angulo}
\end{eqnarray}
where the minus sign in the differential $dy\;$is due to the decrease of the
coordinate $y$.

So the problem of the snowball rolling on an arbitrary trajectory can be
solved by replacing (\ref{Angulo}) into the first of Eqs. (\ref{Ec final}),
and making an assumption like (\ref{Comport masa1}). Additionally, Eq.(\ref%
{Primer ec normal}) provides the solutions for the normal force by
considering Eqs. (\ref{R de curvat}), (\ref{Angulo}) and the solution for
the velocity obtained from (\ref{Ec final}). Notwithstanding, the first of
Eqs. (\ref{Ec final}) does not depend on the normal force whenever the 
\textbf{NSC} is maintained. So, we shall ignore it henceforth.

\section{Comparison of the snowball with other dynamical objects\label%
{sec:comparison}}

May be the clearest way to understand the dynamical behavior of the
snowball, is by comparing it with the dynamics of other simpler objects. In
our case we shall compare the dynamics of four dynamical objects

\begin{enumerate}
\item A skier sliding with no friction (\textbf{SNF}).

\item The same skier but sliding with friction (\textbf{SF}).

\item A ball with constant mass and volume (\textbf{B})

\item A snowball with variable mass and volume (\textbf{SB})
\end{enumerate}

Such comparison will be performed in the $v-t\;$and $v-x\;$planes. The
behavior of the first two objects were reproduced from the article by
Catalfamo \cite{Catalfamo}, and the equations for the ball were obtained
from the ones of the snowball by setting $r\rightarrow constant.\;$In making
the comparison we use the input parameters of table \ref{tab:input}, most of
the parameters in this table were extracted from \cite{Catalfamo}\footnote{%
However, we have changed mildly the parameters that describe the profile of
the exponential trajectory, in order to keep the NSC throughout.}. As it can
be seen, we assume the same mass $M_{0}\;$for the skier and the ball, and
this is also the initial mass of the snowball ending with a mass of $2M_{0}.$

\begin{table}[tbp]
\begin{center}
$%
\begin{tabular}{||c||c||c||c||}
\hline\hline
$M_{0}=K_{0}=85$ & $\rho =917$ & $\rho _{A}=0.9$ & $\mu _{s}=0.03$ \\ 
\hline\hline
$C_{d}=0.3$ & $\theta =\left( 4.76\right) ^{\circ }$ & $A_{S}=0.6$ & $\beta
=0.07$ \\ \hline\hline
$g=9.8$ & $h=25,\ \alpha =0.035$ & $\mu _{D}=0.03$ & $v_{0}=0$ \\ 
\hline\hline
\end{tabular}%
\ $%
\end{center}
\caption{ Input parameters to solve the equation of motion for the \textbf{%
SNF,\ SF, SB,\ }and the \textbf{B.\ }All measurements are in the MKS system
of units.\ $M_{0}\;$is the initial mass of all objects, $K_{0}\;$defines the
increment of mass for the \textbf{SB\ }see eq.(\protect\ref{Comport masa1}%
),\ $\protect\rho \;$and $\protect\rho _{A}\;$are the snow and air densities
respectively, $\protect\mu _{s}\;$the statical coefficient of friction
between the \textbf{B}\ and the ground (and also between the \textbf{SB} and
the ground). $C_{d}\;$is the air drag coefficient,\ $\protect\theta \;$the
angle of the wedge,\ $A_{S}\;$is the skier frontal area, $\protect\beta \;$%
is a parameter that defines the rapidity of increase of mass in the \textbf{%
SB} see Eq. (\protect\ref{Comport masa1}). $g\;$is the gravity acceleration, $%
\protect\mu _{D}\;$is the dynamical coefficient of friction between the 
\textbf{SF\ }and the ground, and $v_{0}\;$is the initial velocity of the
four objects. Further, in the exponential trajectory $y=he^{-\protect\alpha %
x}$, where $h$ is$\;$the height of the hill.}
\label{tab:input}
\end{table}

\subsection{Asymptotic limits\label{sec:asymptotic}}

Before solving all these problems, we shall study the asymptotic limits of
the four objects in the inclined plane and the exponential trajectory with
and without drag force. The asymptotic regime provides useful information
and can be used to analyze the consistency of the numerical solutions. These
limits depend on the drag force, the trajectory and the object itself.

\subsubsection{Inclined plane with no drag force\label{sec:wndf}}

For each object we obtain the following limits

\begin{itemize}
\item For the \textbf{SF}\ its velocity is easily found 
\begin{equation}
v=v_{0}+gt\left( \sin \theta -\mu _{D}\cos \theta \right)  \label{SFND}
\end{equation}
so there is no finite value for the velocity and its behavior is linear
respect to time. The\ \textbf{SNF\ }asymptotic limit\textbf{\ }is obtained
just taking $\mu _{D}\rightarrow 0.$

\item For the \textbf{SB }from Eq.(\ref{Ec final}\textbf{) }and assuming%
\textbf{\ }that the radius reaches an asymptotic limit i.e. $%
dr/dt\rightarrow 0$ when $t\rightarrow \infty \;$we get 
\begin{equation}
v\left( t\rightarrow \infty \right) \equiv v_{\infty }=v_{0}+\frac{5}{7}%
gt\sin \theta  \label{SBND}
\end{equation}%
getting a linear behavior respect to time. The same behavior is exhibited by
the \textbf{B\ }when $t\rightarrow \infty $\textbf{.\ }

Observe that $v_{\infty }\;$in the \textbf{SB} is independent on the mass
and equal to the value for the ball \textbf{B}, it is reasonable owing to
the asymptotic behavior assumed for the \textbf{SB}, Eq. (\ref{Comport masa1}%
).
\end{itemize}

\subsubsection{Inclined plane with drag force}

\begin{itemize}
\item For the \textbf{SF}\ its equation of motion is easily obtained 
\begin{equation*}
\frac{dv}{dt}=-\mu _{D}g\cos \theta -\frac{\rho _{A}C_{d}A}{2M}v^{2}+g\sin
\theta
\end{equation*}
and in the asymptotic limit $dv_{\infty }/dt\rightarrow 0,\;$so we get that 
\begin{equation}
v_{\infty }^{2}=\frac{2g}{\rho _{A}C_{d}}\frac{M}{A}\left( \sin \theta -\mu
_{D}\cos \theta \right)  \label{SF asintot ID}
\end{equation}
Now, by setting $\mu _{D}\rightarrow 0,\;$we get $v_{\infty }\;$for the 
\textbf{SNF}

\item For the \textbf{SB\ }from Eq. (\ref{Ec final}) and setting$%
\;d^{2}s/dt^{2}\rightarrow 0,\;dr/dt\rightarrow 0,\;$we obtain $v_{\infty }$
\end{itemize}

\begin{equation}
v_{\infty }^{2}=\frac{56\rho }{21\rho _{A}C_{d}}r_{\infty }g\sin \theta =%
\frac{2g}{\rho _{A}C_{d}}\frac{M_{\infty }}{A_{\infty }}\sin \theta
\label{SB asintot ID}
\end{equation}%
where $r_{\infty }\equiv r\left( t\rightarrow \infty \right) .\;$The second
term in (\ref{SB asintot ID}) is obtained from Eq.(\ref{Comport masa1}) by
taking the asymptotic limit when $t\rightarrow \infty $, in this case $%
r_{\infty }\;$is given by

\begin{equation*}
r_{\infty }=\left[ \frac{3}{4\pi \rho }\left( M_{0}+K_{0}\right) \right] ^{%
\frac{1}{3}}
\end{equation*}
In the case of the \textbf{B}, we obtain the expression (\ref{SB asintot ID}%
) but $r_{\infty },\;M_{\infty },\;A_{\infty }\;$are constant in time and
equal to its initial values$\;r_{0},\;M_{0},\;A_{0}.$

\subsubsection{Exponential trajectory with no drag force\label{sec:expnodrag}%
}

In this case it is easier to examine the asymptotic limits, since when the
objects have traveled a long path, $\theta \rightarrow 0$ and\ the object is
reduced to run over a horizontal plane, see figure \ref{fig:bola2}.
Therefore the limits are

\begin{itemize}
\item For the \textbf{SF,\ }$v_{\infty }=0.\;$\ 

\item For the \textbf{SNF}, it\textbf{\ }is found easily by energy
conservation 
\begin{equation*}
v_{\infty }^{2}=2gh+v_{0}^{2}
\end{equation*}

where $h\;$is the height of the hill.

\item For the \textbf{B }we can find $v_{\infty }\;$by taking into account
that energy is conserved because friction does not dissipate energy when the 
\textbf{NSC} is held \cite{Finn}. By using energy conservation we obtain%
\footnote{%
Since what really matters for energy conservation is the height of the
center of mass, there is a tiny difference that can be neglected if the
radius of the ball is much smaller that the height of the hill.} 
\begin{equation*}
v_{\infty }^{2}=\frac{10}{7}gh
\end{equation*}%
For the snowball the limit is not easy to obtain because energy is not
conserved and Eq. (\ref{Ec final}) does not provide any useful information%
\footnote{%
By using $\theta \rightarrow 0$,$\;$and $d^{2}s/dt^{2}\rightarrow 0,\;$the
second of Eqs. (\ref{Ec final}) becomes trivial.}. However, according to the
figures \ref{fig:v3}, this limit is lower than the one of the \textbf{B\ }%
owing to the increment of inertia\textbf{.\ }
\end{itemize}

\subsubsection{Exponential trajectory with drag force}

In this case all velocities vanish for sufficiently long time.

\section{Proves of consistency\label{sec:consistency}}

The equations of motion for each object where solved by a fourth order Runge
Kutta integrator\cite{recipes}. To verify the correct implementation of the
program, we reproduce the results obtained by \cite{Catalfamo}, and solve
analitically the problem of the ball of constant radius in the inclined
plane without drag force, the results were compared with the numerical
solution. Additionally, all the asymptotic values discussed above were
obtained consistently.

Finally, the reader could have noticed that one of our main assumptions was
the \textbf{NSC}. However, this condition can only be valid if we ensure
that the statical friction force does not exceed the value$\;\mu _{s}N\;$all
over the trajectory in each case. Otherwise, the snowball would start
slipping preventing us of using the fundamental relations (\ref{NSC1}) and (%
\ref{NSC2}). Additionally, if the snowball started sliding, the frictional
force would become$\;$dynamics i.e. $F_{R_{D}}=\mu _{D}N\;$producing a
coupling among the three Eqs. (\ref{Ec de mov}), remember that we have
assumed that the first one was decoupled. As a first approach, we study the
validity of the \textbf{NSC} in the asymptotic regime by calculating $%
F_{R_{S}}\left( t\rightarrow \infty \right) \;$and $\mu _{s}N\left(
t\rightarrow \infty \right) $.

For the inclined plane, these limits can be estimated by using Eq. (\ref%
{Comport masa1}), the first of Eqs. (\ref{Ec de mov}), and the second of
Eqs. (\ref{Ec final}) 
\begin{eqnarray}
\mu _{s}N\left( t\rightarrow \infty \right) &\rightarrow &\mu _{s}M_{\infty
}g\cos \theta  \notag \\
F_{R_{S}}\left( t\rightarrow \infty \right) &\rightarrow &\frac{2}{5}%
M_{\infty }\left( \frac{d^{2}s}{dt^{2}}\right) _{\infty }.
\label{fricnorasymp}
\end{eqnarray}%
it is easy to verify that the \textbf{NSC\ }is valid in the asymptotic limit
for the wedge, since in the case of the presence of a drag force we get $%
F_{R_{S}}\left( t\rightarrow \infty \right) =0$. Additionally, in the case
of absence of the drag force it is found that $F_{Rs}\left( t\rightarrow
\infty \right) =\left( 2/7\right) \left( M_{0}+K_{0}\right) g\sin \theta ,\;$%
and the condition $F_{R_{S}}\left( t\rightarrow \infty \right) \leq \mu
_{s}N\left( t\rightarrow \infty \right) \;$is accomplished by our input
parameters (see table \ref{tab:input}).

For the exponential trajectory the analysis is even simpler, since the path
for large times becomes a horizontal straight line, the asymptotic limits
for $F_{R_{S}}\;$and $\mu _{s}N$ are the same as in Eqs. (\ref{fricnorasymp}%
) but with $\theta \rightarrow 0$; and the \textbf{NSC}\ condition is held
when$\ t\rightarrow \infty $.

However, the \textbf{NSC} in the asymptotic regime does not guarantee that
it is held throughout the path. For example, in the case of the exponential
trajectory, the maximum slope of the profile is found at the beginning of
the trajectory, it was because of this fact that we changed mildly the
profile parameters defined in Ref. \cite{Catalfamo}. Consequently, by using
the first of Eqs. (\ref{Ec de mov}) as well as the Eqs. (\ref{Ec final}), (%
\ref{Primer ec normal}), (\ref{R de curvat}), and (\ref{Angulo}); we solved
numerically for $F_{R_{S}}\;$vs \emph{time and length} and for $\mu _{s}N$
vs \emph{time and length}; utilizing the numerical input values of table \ref%
{tab:input}. We then checked that $F_{R_{S}}\leq \mu _{s}N\;$throughout the
time or length interval considered in each case.

\section{Analysis\label{sec:analysis}}

In order to make a clear comparison, we take the initial mass of all the
objects to be equal, and all initial velocities are zero. In figures \ref%
{fig:v1}-\ref{fig:v4} we use the following conventions: The dashed dotted
line corresponds to the \textbf{SNF}, the solid line represents the \textbf{%
SF}, the dashed line refers to the \textbf{B},\ and finally the dotted line
corresponds to the \textbf{SB}. In both, ball and snowball we only consider
statical friction and neglect possible dynamic frictional effects due to
sliding, because we have guaranteed that the \textbf{NSC\ }is valid
throughout the trajectory, as explained in section \ref{sec:consistency}.

\begin{center}
\begin{figure}[tbh]
\resizebox{\textwidth}{!}{
\rotatebox{0}{\includegraphics[bb=0 0 800 600]{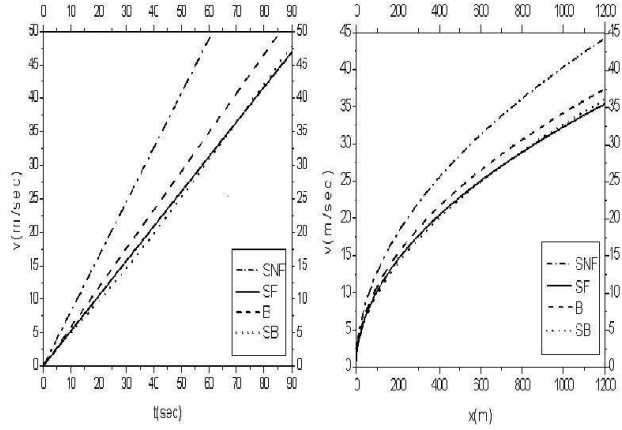}}}
\caption{\textit{Plots in the $v-t\;$plane (left) and the $v-x\;$plane
(right) when the objects travel in a wedge of constant slope with no drag
force.}}
\label{fig:v1}
\end{figure}

\begin{figure}[tbh]
\resizebox{\textwidth}{!}{
\rotatebox{0}{\includegraphics[bb=0 0 800 600]{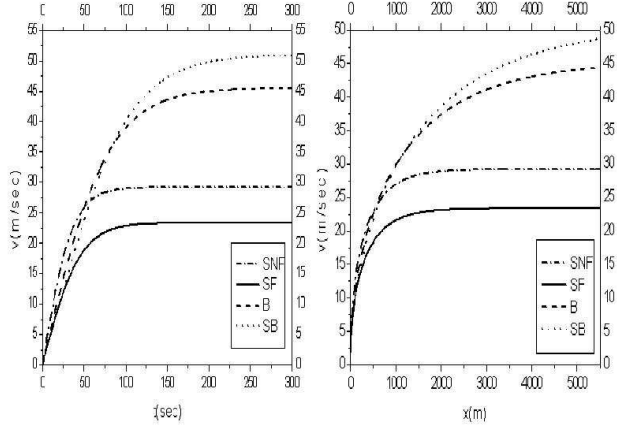}}}
\caption{\textit{Plots in the $v-t\;$plane (left) and the $v-x\;$plane
(right)\ when the objects run over a wedge of constant slope with drag force.%
}}
\label{fig:v2}
\end{figure}

\begin{figure}[tbh]
\resizebox{\textwidth}{!}{
\rotatebox{0}{\includegraphics[bb=0 0 800 600]{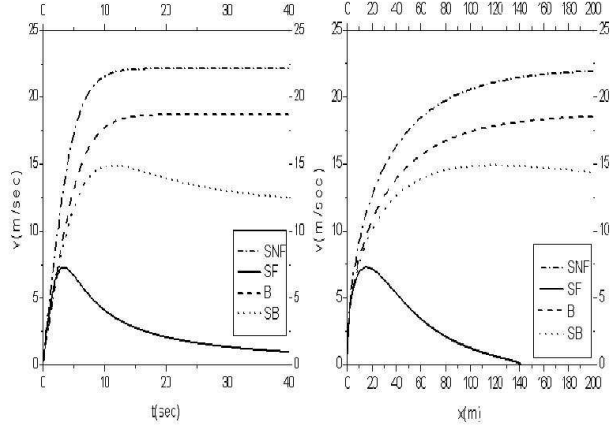}}}
\caption{\textit{Plots in the $v-t\;$plane (left) and the $v-x\;$plane
(right) when the objects travel on an exponential trajectory with no drag
force.}}
\label{fig:v3}
\end{figure}

\begin{figure}[tbh]
\resizebox{\textwidth}{!}{
\rotatebox{0}{\includegraphics[bb=0 0 800 600]{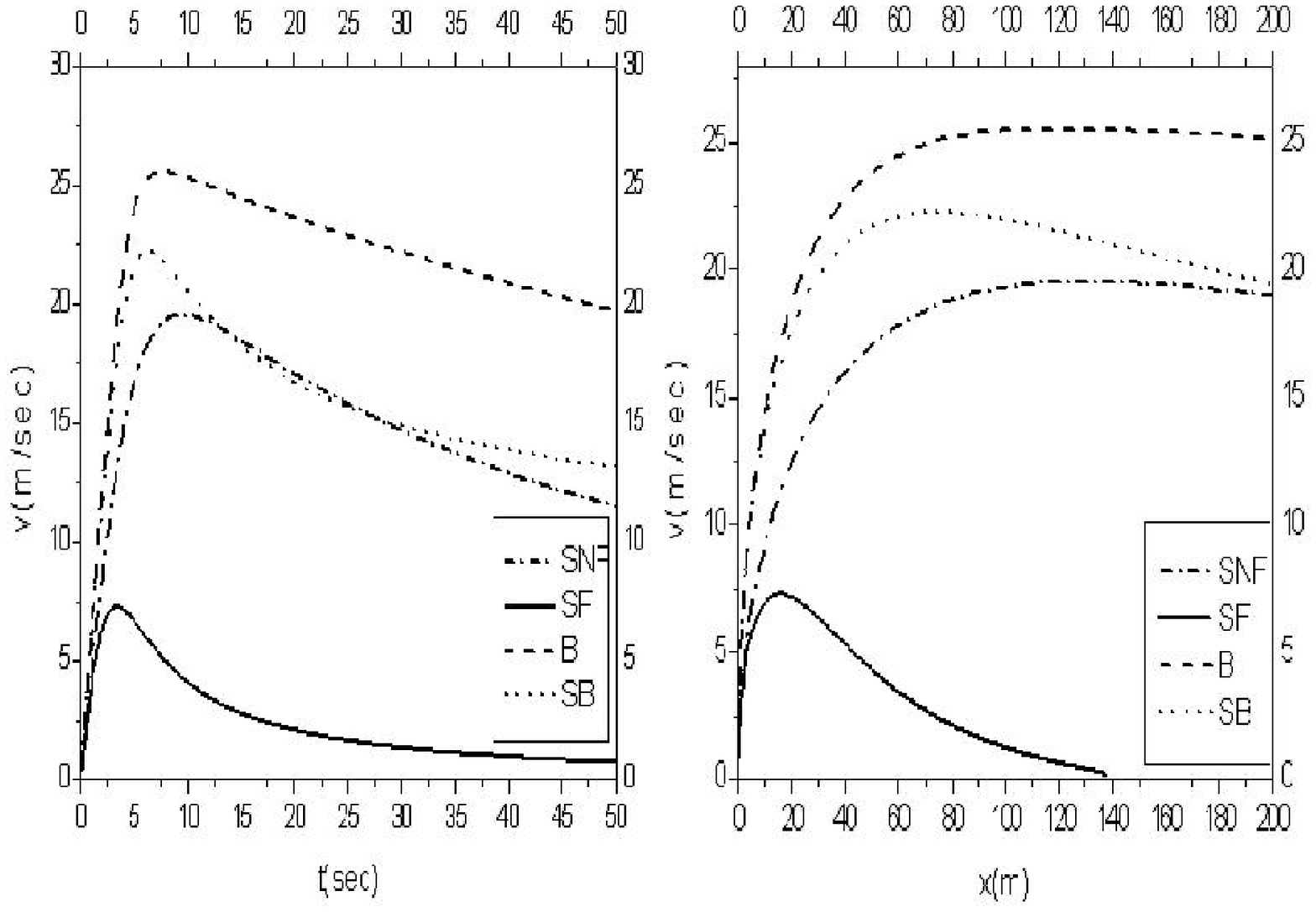}}}
\caption{\textit{Plot in the $v-t\;$plane (left) and the $v-x\;$plane
(right) when the objects run over an exponential trajectory with drag force.}
\textit{All velocities vanish for sufficiently long time (or length) as
expected, despite it is not shown in the interval displayed.}}
\label{fig:v4}
\end{figure}
\end{center}

\textbf{In figure \ref{fig:v1}} we plot $v\;$vs\ $t\;$and $v\;$vs $x\;$for
constant slope of the wedge without drag force. Of course, all graphics in
the $v-t\;$plane are straigth lines except the one for the snowball. We can
see that the \textbf{SNF} is the fastest object as expected, since no
retardation factors are acting on it, next we have the \textbf{B}\ which
posseses the rotation as a retardation factor. Additionally, the \textbf{SB}
line is always below the \textbf{B} line because in the former, two
retardation factors are present: the rotation and the increase of inertia.
However, for sufficiently long time (or length) the increase of inertia
vanishes (according to our assumptions) so that the velocities of both 
\textbf{B} and \textbf{SB} coincide, in agreement with the analysis made in
Sec. \ref{sec:wndf}. We checked that condition, though it does not appear in
the Fig. \ref{fig:v1}, because of the short time and length interval
displayed.

The line corresponding to the \textbf{SF} is below the line corresponding to
the \textbf{SNF} as it must be, however the relation between the \textbf{SF}
and \textbf{SB} lines is particularly interesting and deserves more
attention. At the beginning the \textbf{SB} is slightly slower than the 
\textbf{SF,} but for sufficiently long time, the \textbf{SB} becomes faster.
It can be explained in the following way, at the beginning the \textbf{SB}
has two retardation factors: the rotation and the increase of inertia, while
the \textbf{SF} only has one retardation factor: the sliding friction. On
the other hand, for sufficiently long time the increment of inertia becomes
negligible in the\ \textbf{SB}, and only the rotation acts as a retardation
factor, consequently the \textbf{SB} behaves like the \textbf{B} as shown in
section \ref{sec:comparison}. Therefore, the combination of the two
retardation factors at the beginning makes the \textbf{SB} slower but when
the increase of inertia is small enough, the \textbf{SB} becomes faster than
the \textbf{SF}. Nevertheless, we should point out that this behavior depend
on the value of $\mu _{D}$,$\;$if it were large enough the line for the 
\textbf{SF} would lie below the lines for \textbf{B}\ and \textbf{SB} at all
times, in contrast if it were small enough the \textbf{SF} line would lie
above the \textbf{B} and \textbf{SB} lines. Notwithstanding, the rapidity of
the \textbf{SF} must be smaller than the \textbf{SNF} speed at any time and
for any value of $\mu _{D}.\;$

According to this analysis, when the objects travel in a wedge with no drag
force, the pattern of velocities in descendent order for any set of the
input parameters (as long as the initial masses and velocities are the same)
is the following: the \textbf{SNF}, the \textbf{B} and the \textbf{SB}. The
comparative velocity of the \textbf{SF} depend on the input parameters but
it must be always slower than the \textbf{SNF}. As a proof of consistency,
it can be checked that the asymptotic limits in Eqs. (\ref{SFND}), (\ref%
{SBND}) obey this pattern.

\textbf{Figure \ref{fig:v2}} correspond to a wedge with constant slope
including drag force. In this case the comparative behavior among the four
elements is not as simple as in figure \ref{fig:v1}, because in this case
the lines cross out each other. However, the line describing the \textbf{SF}
is always below the line describing the \textbf{SNF} as it should be. This
more complex behaviour owes to the frontal area dependence of the drag
force. For instance, we can realize that at short times the comparative
behavior is very similar to the one in figure \ref{fig:v1}, since the drag
force has not still acted significantly. All these elements get an
asymptotic limit as we described in section \ref{sec:comparison}. We see
that the largest asymptotic limit correspond to the \textbf{SB}, in
opposition to the case of figure \ref{fig:v1} with no drag force, in which
the snowball was one of the slowest objects; the clue to explain this fact
recides in the frontal area dependence of the drag force. From Eqs. (\ref{SF
asintot ID}, \ref{SB asintot ID}) we can verify that for all these objects
the terminal velocity behaves as $v^{2}\varpropto M/A,\;$this quotient is
larger for the \textbf{B} than for the \textbf{SNF} and the \textbf{SF} in
our case, then the asymptotic velocity $v_{B}\;$is larger than $v_{SNF}\;$%
and $v_{SF}$,\ for both the skier and the ball this ratio is a constant. In
contrast, since in the snowball the mass grows cubically with the radius
while the area grows quadratically, its velocity behaves such that $%
v_{SB}^{2}\varpropto r\left( t\right) .\;$Therefore, for sufficiently long
times, its velocity grows with the radius of the \textbf{SB}, getting a
higher terminal velocity (of course it depends on the asymptotic value of $%
r\left( t\right) $). Observe that if we had assumed a non asymptotic
behavior of $r\left( t\right) \;$in (\ref{Ec final}) we would have not
obtained any finite terminal velocity for the snowball even in the presence
of a drag force. Furthermore, we see that the terminal velocity for the 
\textbf{SB\ }is reached in a longer time than the others, it is also because
of the slow growth of $r(t).$

\textbf{Figure \ref{fig:v3}} describes the four elements traveling in an
exponential hill with no drag force. Two features deserve special attention:
(1) the terminal velocity is achieved in a very short time specially in the
cases of the \textbf{SNF} and the \textbf{B}, these limits coincides with
the ones obtained in section \ref{sec:comparison}. (2) For the \textbf{SB}
and the \textbf{SF} there is a local maximum velocity at a rather short
time, the diminution in the velocity since then on, owes to the decreasing
in the slope of the path in both cases, the increment of inertia in the case
of the \textbf{SB}, and the friction in the \textbf{SF}. Such local maximal
velocity cannot be exhibited by the \textbf{SNF} and the \textbf{B} because
conservation of energy applies for them, and as they are always descending
their velocities are always increasing, though for long times they are
practically at the same height henceforth, getting the terminal velocity. In
particular, we see that the terminal velocity of the \textbf{SF} is zero as
it was shown in Sec. \ref{sec:expnodrag}.

\textbf{In figure \ref{fig:v4}} the elements travel in an exponential hill
with drag force. In this case, the conservation of energy does not hold for
any of the objects, consequently maximum velocities in intermediate steps of
the trajectory are allowed for all of them. All terminal velocities are zero
as expected. Because of the same arguments discussed above, the line of the 
\textbf{SF} is below to the one of the \textbf{SNF}. However, any other
pattern depend on the numerical factors utilized.

A final comment is in order, we can realize that though the solution of the
kinetics of the snowball depends on the ansatz made about the mass growth,
the bulk of our results and analysis only depend on the fact that the
snowball mass reaches a finite asymptotic value. So that the discussion is
quite general, especially in the asymptotic regime.

\section{Conclusions\label{sec:conclusions}}

We have described the behavior of a snowball acquiring mass while rolling
downhill, taking into account the enviromental conditions. The dynamics of
the snowball is very complex because it is a rotating object and at the same
time its mass and moment of inertia are variables. In order to visualize
better the effects due to the rotation and mass variation, we compare its
motion with the kinetics of two objects in which the rotation and mass
variational effects are absent (the \textbf{S}kier with \textbf{F}riction
and the \textbf{S}kier with \textbf{N}o \textbf{F}riction), and with one
object in which the rotation is present but no the mass variation (the 
\textbf{B}all of constant mass and radius).

The comparative behavior of these objects depend on the trajectory but also
on some retardation factors: the friction, the drag force, the increase of
mass (inertia), and the rotational effects. It worths to remark that despite
the increment of inertia is a retardation factor in some circumstances, it
could surprisingly diminish the retardation effect due to the drag force. In
addition, some local maxima of the velocities for each object appears in an
exponential trajectory, showing that the maximum velocity might be achieved
at an intermediate step of the path.

Finally, we point out that despite the complete solution of the snowball
depends on an ansatz about the way in which its mass grows; its comparative
dynamics respect to the other objects in the asymptotic regime is basically
independent of the details of the growth, and only depend on the assumption
that the mass reaches an asymptotic value, a very reasonable supposition.
Therefore, we consider that our analysis is not very model dependent at
least in the regime of large times or lengths. In addition, these asymptotic
limits serves also to show the consistency of our results.

\end{document}